\documentclass[conference]{IEEEtran}
\IEEEoverridecommandlockouts
\usepackage{cite}
\usepackage{amsmath,amssymb,amsfonts}
\usepackage{algorithmic}
\usepackage{graphicx}
\usepackage{textcomp}
\usepackage{xcolor}

\usepackage{enumitem}
\usepackage{enumerate}
\usepackage{listings,jvlisting}

\usepackage{inconsolata}

\usepackage{multirow}

\usepackage{tikz} 
\usepackage{lipsum} 
\makeatletter 

\def\ieeecopyright{ \footnotesize © 2025 IEEE. Personal use of this material is permitted.\newline DOI: 10.1109/ISORC65339.2025.00051} \makeatother \AddToHook{shipout/firstpage}{
\begin{tikzpicture}[remember picture,overlay] \node[anchor=south west,xshift=1.0cm,yshift=0.8cm] at (current page.south west){\parbox{\linewidth}{\raggedright\ieeecopyright}}; \end{tikzpicture} }

\def\BibTeX{{\rm B\kern-.05em{\sc i\kern-.025em b}\kern-.08em
    T\kern-.1667em\lower.7ex\hbox{E}\kern-.125emX}}
\begin{document}

\title{TECS/Rust-OE: Optimizing Exclusive Control in Rust-based Component Systems for Embedded Devices
}

\author{\IEEEauthorblockN{Nao Yoshimura}
\IEEEauthorblockA{\textit{Graduate School of}\\
\textit{Science and Engineering}\\
\textit{Saitama University}\\
}
\and
\IEEEauthorblockN{Hiroshi Oyama}
\IEEEauthorblockA{\textit{OKUMA Corporation}}
\and
\IEEEauthorblockN{Takuya Azumi}
\IEEEauthorblockA{\textit{Graduate School of}\\
    \textit{Science and Engineering}\\
    \textit{Saitama University}
}

}

\maketitle

\begin{abstract}

The diversification of functionalities and the development of the IoT are making embedded systems larger and more complex in structure. Ensuring system reliability, especially in terms of security, necessitates selecting an appropriate programming language. As part of existing research, TECS/Rust has been proposed as a framework that combines Rust and component-based development (CBD) to enable scalable system design and enhanced reliability. This framework represents system structures using static mutable variables, but excessive exclusive controls applied to ensure thread safety have led to performance degradation. This paper proposes TECS/Rust-OE, a memory-safe CBD framework utilizing call flows to address these limitations. The proposed Rust code leverages  real-time OS exclusive control mechanisms, optimizing performance without compromising reusability. Rust code is automatically generated based on component descriptions. Evaluations demonstrate reduced overhead due to optimized exclusion control and high reusability of the generated code.

\end{abstract}


\vspace{-9.7pt}

\section{Introduction}

\vspace{-1.7pt}

Embedded systems are widely used in a variety of fields, including IoT devices, in-vehicle systems, and medical equipment. However, these systems face challenges of increasing scale and complexity~\cite{FMP3+TECS, LargeScaleAndComplex}, and security is becoming more difficult to ensure. Embedded systems are developed in C and C++ due to their real-time nature, but memory management and safety of these languages are dependent on the developer, leading to memory-related bugs. To ensure memory safety, a programming language called Rust is being adopted for embedded systems development. Rust guarantees memory safety at compile time, allowing the development of highly secure software.

For embedded systems, which are becoming larger and more complex, a framework combining Rust and \textit{component-based development} (CBD)~\cite{TECS_ISORC2010, ThreeLayerCBD} was proposed~\cite{TECSRust}. This framework enhances system scalability and reliability while enabling efficient development in resource-constrained environments. Although this framework builds up static software structures, static mutable instances are required to be secured by Rust. Therefore, all static mutable instances are subject to exclusive control, resulting in unnecessary overhead.

To address these issues, this paper proposes TECS/Rust-OE, a memory-safe component framework for embedded systems, focusing on optimizing exclusive control and focuses on \textit{TOPPERS Embedded Component Systems} (TECS)~\cite{TECS}, a component framework for embedded systems. The proposed framework optimizes the exclusive controls while maintaining high code reusability. Optimization determines the need for exclusive controls, reduces the number, and selects the appropriate exclusive control depending on the situation. Code reflecting the optimization is automatically generated to support efficient development.

This paper is organized as follows. Section II discusses the system model and the assumptions of the embedded component system in this study. Section III describes the development flow, design, and implementation. Section IV presents the evaluation results. Section V introduces and compares related studies. Section VI provides the conclusion of this study.

\vspace{-9.7pt}
\section{System Model}

\vspace{-1.7pt} 

\begin{figure}[t]
\begin{center}
        \includegraphics[width=1\linewidth]{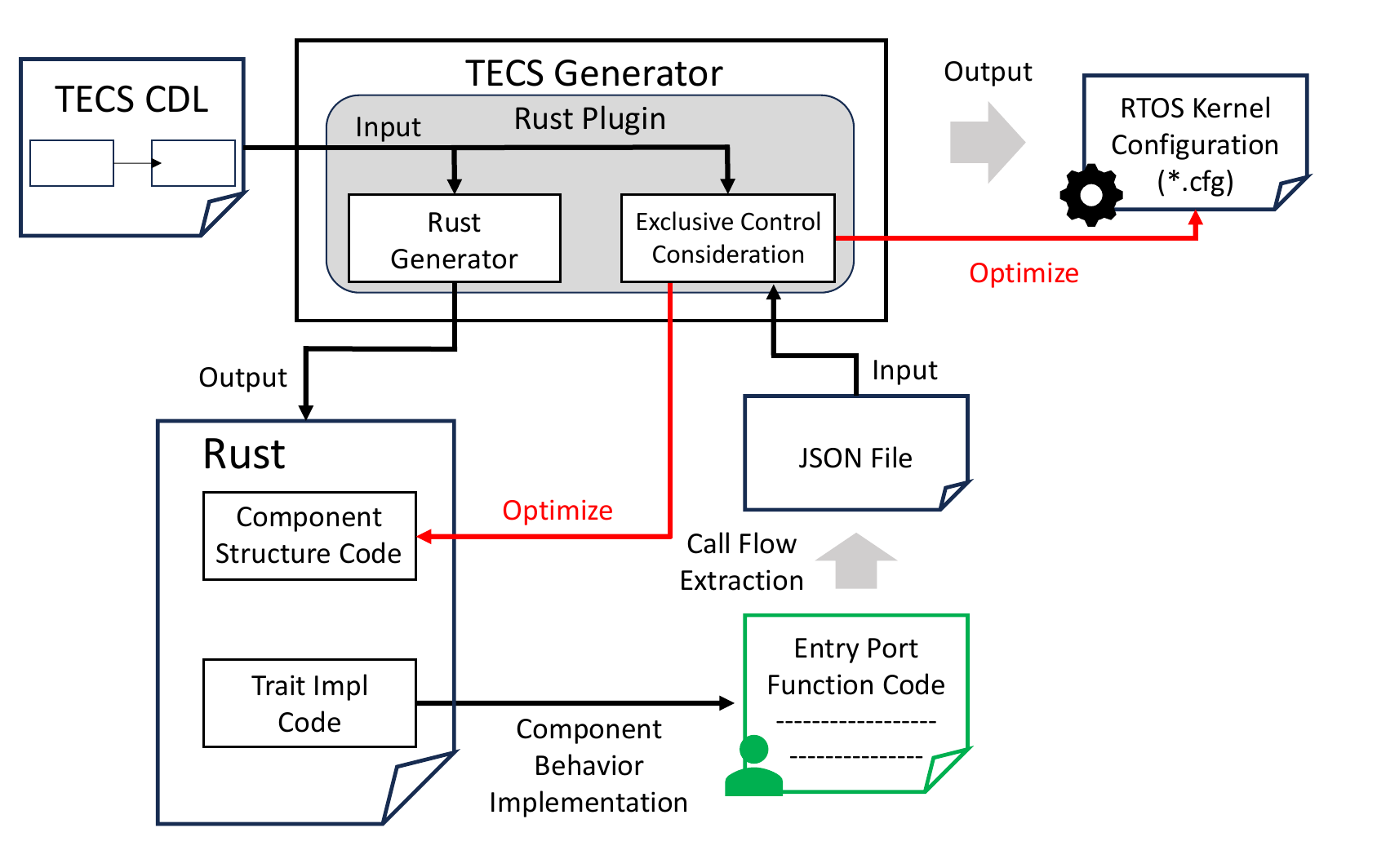}
        \vspace{-10.7pt}
	\caption{System model.}
    \label{fig:System model}
    \vspace{-23.7pt}
\end{center}
\end{figure}

The system model of the proposed framework, TECS, and Rust are described in this section. The system model structure of the proposed framework is shown in Fig.~\ref{fig:System model}. The proposed framework optimizes the Rust code that represents the component structure based on the coupling relationship of components and call flow. In the first operation, the TECS generator generates Rust code from the component description file. Part of the generated Rust code is used by the developer as \textit{Entry Port Function Code} to implement the behavior of the component. The call flow extracted from the \textit{Entry Port Function Code} is used to run the TECS generator for the second time, optimizing the generated Rust code and generating the RTOS configuration file accordingly. 

\subsection{TECS}

TECS is a component framework for embedded systems that improves software reusability and visualization. A component is a piece of software into which software is divided, and a system is composed of these components statically combined. Componentization improves system maintainability by separating the interface from the implementation and facilitating implementation changes.

In TECS, component definitions and coupling relationships are described in \textit{Component Description Language} (CDL). The type of a component is called a \textit{celltype}, which includes \textit{entry ports}, \textit{call ports}, \textit{attributes} that are constant, and static mutable \textit{variables}. An \textit{entry port} provides the functionality of its \textit{celltype}, and a \textit{call port} is a port that uses the functionality of other components. The set of functions that interface between components is called a \textit{signature}, and is required when defining the ports of a \textit{celltype}. An instance of a \textit{celltype} is called a \textit{cell} and represents how the components are connected.

\vspace{-2.7pt}
\subsection{Rust}

Rust~\cite{Rust} is designed with a focus on memory safety, parallelism, and speed. The memory safety is achieved through concepts such as \textit{ownership} systems, \textit{borrowing} systems, and lifetime annotation. These concepts allow the Rust compiler to point out memory-related bugs such as incorrect memory references and dangling pointers.

An interface in Rust is called a \textit{trait} and defines behavior for a type. Specifying a \textit{trait} to generics is called a \textit{trait bound}, which guarantees that the generics implement the \textit{trait}. A \textit{trait} is also treated as an object, abstracting the type that implements the \textit{trait}. This abstraction allows for dynamic dispatching, where the actual method to be invoked is determined at runtime.

\vspace{-5.7pt}
\section{Proposed Framework}
\label{ssub: Proposed Framework}

\begin{figure}[t]
\begin{center} 
        \includegraphics[width=1\linewidth]{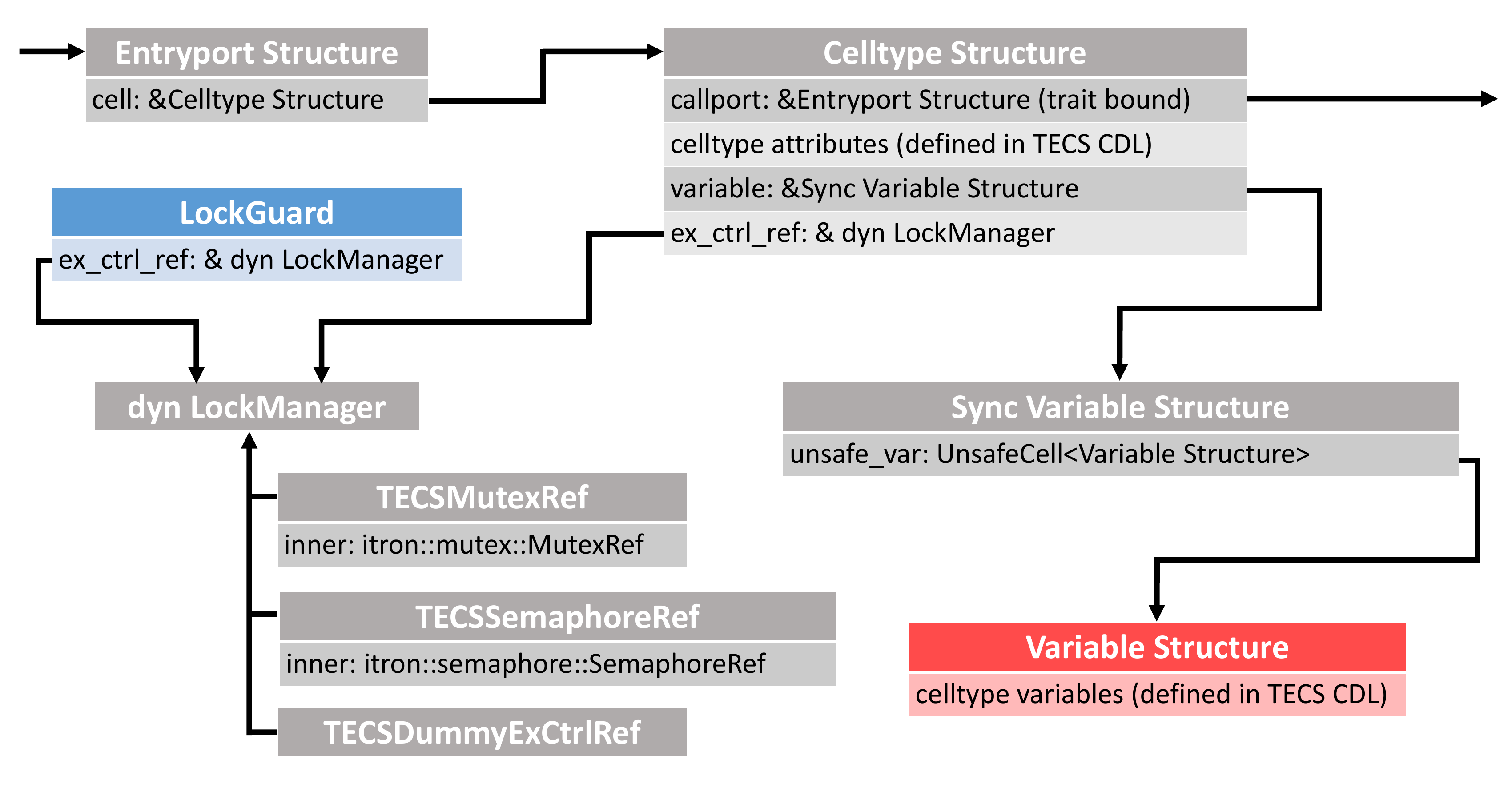}
        \vspace{-20.7pt}
	\caption{Data structure of Rust code considering exclusive control.}
    \label{fig: Data structure of Rust code considering exclusive control}
\end{center}
\vspace{-17.7pt}
\end{figure}

    \vspace{-1.7pt}
    \subsection{Optimization Flow and Generated Rust Code}
    \label{ssub: Optimization flow and generated Rust code}

    In this section, the flow of optimization of exclusive control and the data structure of Rust code for optimization are introduced. Exclusive control optimization requires the component coupling relationship information in the CDL file and the component call flow in the \textit{Entry Port Function Code}. This information is extracted by existing tools and their extensions and used to generate Rust code automatically. In addition, data structures not presented in this paper follow the TECS/Rust v1.

        \subsubsection{Data Structure of Rust Code Considering Exclusive Control}
        \label{ssub: Data Structure of Rust Code Considering Exclusive Control}

        The data structure of the Rust code, which is compatible with CDL and takes into account exclusive control, is described. This is achieved by modifying the already proposed TECS/Rust~\cite{TECSRust} code to optimize exclusive control while maintaining high reusability.

        The data structure corresponding to a \textit{celltype} in CDL is shown in Fig.~\ref{fig: Data structure of Rust code considering exclusive control} and is called a \textit{celltype structure} and is instantiated statically. The \textit{celltype structure} has a \textit{call port} and \textit{attributes} defined in the CDL. The type of the \textit{call port} field is set to the \textit{trait bound} corresponding to the connected \textit{signature}. \textit{Variables} of \textit{celltype} in CDL correspond to the variable field of \textit{celltype structure}, but the data structure is different from \textit{call port} and \textit{attributes}.

        \textit{Variables} of \textit{celltype} are defined using \textit{sync variable structure}, UnsafeCell, and \textit{variable structure}. The \textit{variable structure} contains the entity of the \textit{variables} defined in the CDL, wrapped by the UnsafeCell and instantiated into a static mutable. The wrapped data is owned by the unsafe\_var field of the \textit{sync variable structure}. The unsafe\_var field is in a state where the restrictions of Rust are relaxed by UnsafeCell.

        UnsafeCell is a type that relaxes the sharing and variability limitations of Rust and provides internal mutability. The \textit{sync variable structure} has a \textit{variable structure} that is mutable, but the \textit{sync variable structure} itself is referenced immutably by the \textit{celltype structure}. This is the state of being externally immutably yet internally mutable. This state violates the rules of Rust and cannot be compiled. Therefore, UnsafeCell is used to enable compilation. However, UnsafeCell is not thread-safe in this data structure, making the \textit{variable structure} unsafe.

        \subsubsection{Variable Structure Safety Assurance}
        \label{ssub: Variable Structure Safety Assurance}

        Thread-safety of the \textit{variable structure} is guaranteed by the ex\_ctrl\_ref field of the \textit{celltype structure}. The ex\_ctrl\_ref field is a reference to a custom exclusive control type that wraps the ITRON-specification RTOS exclusive control mechanism. The custom exclusive control type TECSMutexRef has a reference to a mutex object, and TECSSemaphoreRef has a reference to a semaphore object. The inner field of each type uses the types provided by itron crate, one of the Rust libraries. The custom exclusive control types are also abstracted by the implementation of LockManager, a \textit{trait} that provides exclusive control locking and unlocking operations.

\begin{figure}[t]
\centering
	\begin{lstlisting}
 impl SSensor for ESensorForTSensor<'_>{
   #[inline]
   fn set_device_ref(&'static self) {
     let (port, var, _lg) = self.cell.get_cell_ref();
     // Developers implement the component behavior here.
   }
   // Other fucntions
	\end{lstlisting}
\vspace{-5.7pt}
\caption{Example of trait impl code.}
\label{fig: Impl file code}
\vspace{-10.7pt}
\end{figure}

        The exclusive control operations for the \textit{variable structure} utilize the get\_cell\_ref function shown in Fig.~\ref{fig: Impl file code} and LockGuard shown in Fig.~\ref{fig: Data structure of Rust code considering exclusive control}. get\_cell\_ref is called first when the developer implements the \textit{Entry Port Function Code}. Exclusive control is initiated by locking for a \textit{variable structure} in this function. This function returns a LockGuard in addition to references to the \textit{call ports}, \textit{attributes}, and \textit{variables} of the celltype. Therefore, the lifetime of LockGuard is the same as the scope of the call to get\_cell\_ref. LockGuard implements a destructor that unlocks the exclusive control and automatically unlocks it when the lifetime runs out. The interval of the exclusive control is equal to the lifetime of the functionality provided by the \textit{entry port}, and the safety of the \textit{variable structure} is guaranteed.

        \subsubsection{Exclusive Control Optimization}
        \label{ssub: Exclusive Control Optimization}

        The optimization of the exclusive control is achieved by running the TECS generator twice. The first run generates code with exclusive control applied to all cells with \textit{variables}. In the second run, the call flow is used to generate the optimized code and the corresponding RTOS configuration file.

        The need for exclusive control is determined from the coupling relationship of the components and the call flow. From the information in the call flow, check from which task each cell is accessed. Cells accessed only by a single task do not require exclusive control. Cells accessed by multiple tasks require exclusive control. However, if cells accessed by the same task are combined in a sequence, the exclusive controls are removed, except for the first cell.

        The removal of exclusive control is achieved by utilizing dummies and changing the definition of the \textit{celltype structure}. The dummy is TECSDummyExCtrlRef in Fig.~\ref{fig: Data structure of Rust code considering exclusive control}, which is a structure with no fields. This dummy is abstracted by LockManager, but the implementation of locking and unlocking is empty.

        Dummy is used to represent a \textit{celltype} with a mixture of cells that require exclusive control and cells that do not. The structure in Fig.~\ref{fig: Data structure of Rust code considering exclusive control} represents this case, where the ex\_ctrl\_ref field of the \textit{celltype structure} is the LockManager Trait object. This abstraction of the \textit{trait} object allows the use of dummies only for cells that do not require exclusive control.

        The \textit{celltype} with only cells that do not require exclusive control is represented by deleting the ex\_ctrl\_ref field. In addition, the \textit{celltype} with only cells that need exclusive control is represented by making the reference of the ex\_ctrl\_ref field in Fig.~\ref{fig: Data structure of Rust code considering exclusive control} a direct reference to TECSMutexRef or TECSSemaphoreRef. The choice of mutex and semaphore is determined based on the information of the tasks accessing the cell. If more than three tasks are to be accessed and if they have different priorities, the exclusive control is a mutex employing the priority ceiling protocol; otherwise, a semaphore.

\vspace{-7.7pt}
\section{Evaluation}
\label{ssub: Evaluation}

This section compares Rust and the proposed framework and evaluates the proposed framework. To evaluate the performance improvement of the optimization with the proposed framework, the execution time is measured. In addition, the number of lines of code is measured to evaluate the reusability of the auto-generated code.
The RTOS used for the measurements is TOPPERS/ASP3. The execution time is measured by running TOPPERS/ASP3 on STM32F413VG.

    \vspace{-3.7pt}
    \subsection{Comparison with Rust Execution Time}
    \label{ssub: Comparison with Rust Execution Time}

    This section compares the execution times of ordinary Rust and Rust using TECS. The comparison uses APIs for motors and sensors in SPIKE-RT~\cite{SPIKE-RT}. The three APIs used for comparison are get\_distance, stop, and set\_speed. The comparison confirms the overhead and changes caused by the proposed approach.

\begin{figure}[t]
\begin{center} 
        \includegraphics[width=0.97\linewidth]{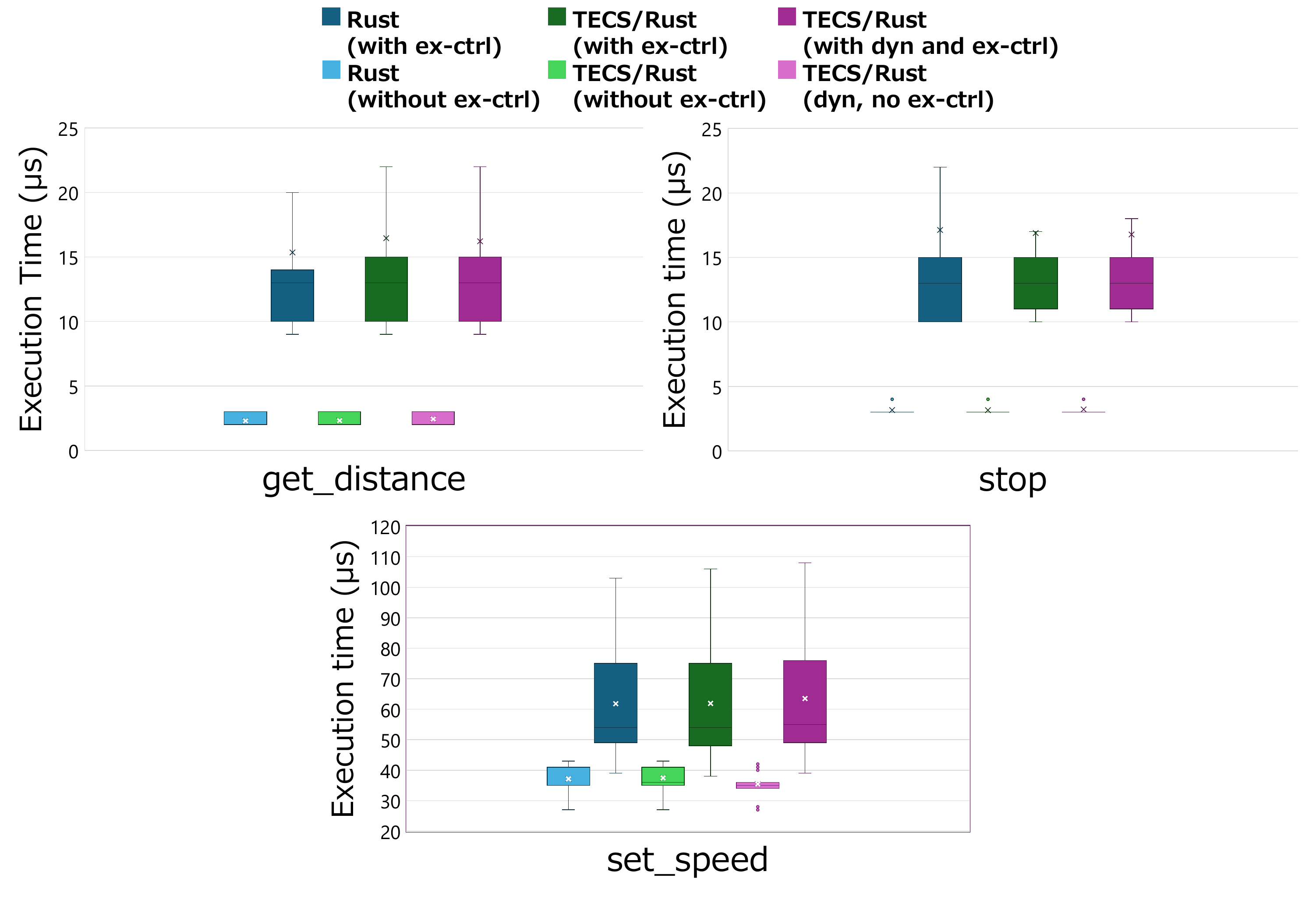}
        \vspace{-4.7pt}
	\caption{Evaluation Time of SPIKE-RT APIs.}
    \label{fig: Evaluation Time}
\end{center}
\vspace{-17.7pt}
\end{figure}

    Execution time results are shown in Fig.~\ref{fig: Evaluation Time}, with Rust results in light blue and blue. The other graphs show the results of the proposed framework. The results in blue, green, and purple are the results where the TOPPERS exclusive control is applied. This exclusive control is a mutex object of the priority ceiling protocol. 

    The measurement results show no difference between the proposed framework and ordinary Rust. This indicates that no overhead is incurred due to componentization. The light purple and purple graphs show the results of the dynamic dispatching described in Section~\ref{ssub: Exclusive Control Optimization} applied to the proposed framework. These results show that dynamic dispatching does not affect the execution time, and that proper invocation of exclusive control operations at runtime is achieved. The large difference in execution time between the light green and green graphs is the execution time of operations related to exclusive control. The removal of exclusive control through optimization reduces overhead and improves predictability. The reduction effect is expected to be more pronounced in systems where the same component is used many times. Therefore, the proposed framework eliminates unnecessary exclusive control and reduces overhead while achieving system componentization.

    \vspace{-2.7pt}
    \subsection{Number of Lines of Code Generated and Comparison Before and After Optimization}
    \label{ssub: Number of Lines of Code Generated and Comparison Before and After Optimization}

\begin{table}[t]
    \caption{Comparison of the number of lines of code}
    \label{tab: Comparison of the number of lines of code}
    \centering
    \scalebox{0.97}{
    {\tabcolsep = 1.25mm
    \begin{tabular}{|l|c|c|c|c|}
    \hline
        & \multirow{2}{*}{Rust} & \multicolumn{2}{c|}{TECS/Rust}\\
        \cline{3-4}
        & & Before optimize & After optimize \\
        \hline
        CDL File & 0 & 24 & 0 \\
        \hline
        Auto-generated Code & 0 & 629 & 588 \\
        \hline
        Written Code & 237 & 196 & 0 \\
        \hline
        Hand-coding & 237 & 220 & 0 \\
        \hline
        Compiled Code & 237 & 825 & 784 \\
        \hline
    \end{tabular}
    }
    }
\vspace{-10.7pt}
\end{table}

    In this section, the number of lines of code generated by the TECS generator and the amount of hand coding were measured and the results are shown in Table~\ref{tab: Comparison of the number of lines of code}. Checking the number of lines of code before and after optimization reveals the impact of the generated code on the \textit{Entry port Function Code}. The code used for the measurement is a simple application for SPIKE-RT, which controls a motor according to the distance acquired from an ultrasonic sensor.

    Hand-coding is the value of all lines the user needs to write. The TECS/Rust value must take into account the number of lines in the CDL, which is the sum of the CDL File and the Written Code value. Comparing the Hand-coding values of TECS/Rust and normal Rust, TECS/Rust is smaller. This is because the \textit{celltype} description of SPIKE-RT is reused and the CDL File value is smaller. By reusing the \textit{Entry port Function Code} in the Written Code, the value of Hand-coding can be made closer to zero in the future. This indicates that the reusability of the proposed framework makes system development more efficient.

    The Hand-coding values before and after optimization indicate high reusability of the proposed Rust code. The value of Hand-coding after optimization of TECS/Rust is a perfect zero. This is because the scope of influence of the optimization is limited to Auto-generated Code. This value of Hand-coding indicates that the user does not need to make any changes to the code due to exclusive control optimization. This smoothes the process from initial code generation to optimization, supporting smooth development while preserving reusability of the code.

\vspace{-10.7pt}
\section{Related Work}

In this chapter, existing studies involving CBD are presented and compared with the proposed framework.

    CRC is a framework that combines the memory model of Rust with reactive components~\cite{CRC}. CRC focuses on single-core and provides priority-aware scheduling that does not generate deadlocks. Static analysis of system structure enables automatic generation of memory-efficient Rust code.

    ASP3+TECS is one of the studies related to TECS, which componentizes TOPPERS/ASP3 by TECS~\cite{ASP3+TECS}. ASP3 is a standard profile among the TOPPERS RTOSs. ASP3+TECS treats kernel objects as components and enables configuration simply by describing the system structure.

    HRMP3+TECS is one of the TECS-related studies, in which TOPPERS/HRMP3 is componentized by TECS~\cite{HRMP3+TECS, HRMP3+TECSv2}. HRMP3 provides memory protection and time partitioning functionality for multi-core systems to achieve more reliable systems. These functionality can be setup from TECS and enable efficient configuration.

    ASP3+TECS supports efficient development by components, but their memory safety must be the responsibility of the developer. HRMP3+TECS provides memory protection, but memory protection is a different concept from memory safety. Unlike these frameworks, the proposed framework provides reliable development because memory safety is guaranteed by Rust.

    CRC is the same as the proposed framework in terms of providing memory-safe CBD. However, CRC focuses on a single-core bare-metal environment. The proposed framework is easier to integrate with the RTOS and can be used for multi-core environments, which is a clear difference.

\vspace{-3.7pt}
\section{Conclusion}

This paper proposed a framework to improve the performance of memory-safe CBD for embedded systems while maintaining its reliability. The proposed framework utilized call flow to determine the need for exclusive control of components. The Rust code was determined based on the need for exclusive control, and optimization was achieved while maintaining high reusability. The component structure of TECS guarantees the safety of Rust in the parts where exclusive control is removed by optimization. The performance improvement due to optimization and the selection of appropriate exclusive control was demonstrated by measuring the execution time. The reusability of the proposed code was demonstrated by measuring the number of lines of code generated and the amount of hand coding.

\vspace{-3.7pt}
\section*{Acknowledgment}
This work was supported by JSPS KAKENHI Grant Number 23H00464.

\vspace{-3.7pt}

\bibliographystyle{IEEEtran}
\bibliography{bibtex}

\begin{thebibliography}{10}
\providecommand{\url}[1]{#1}
\csname url@samestyle\endcsname
\providecommand{\newblock}{\relax}
\providecommand{\bibinfo}[2]{#2}
\providecommand{\BIBentrySTDinterwordspacing}{\spaceskip=0pt\relax}
\providecommand{\BIBentryALTinterwordstretchfactor}{4}
\providecommand{\BIBentryALTinterwordspacing}{\spaceskip=\fontdimen2\font plus
\BIBentryALTinterwordstretchfactor\fontdimen3\font minus \fontdimen4\font\relax}
\providecommand{\BIBforeignlanguage}[2]{{%
\expandafter\ifx\csname l@#1\endcsname\relax
\typeout{** WARNING: IEEEtran.bst: No hyphenation pattern has been}%
\typeout{** loaded for the language `#1'. Using the pattern for}%
\typeout{** the default language instead.}%
\else
\language=\csname l@#1\endcsname
\fi
#2}}
\providecommand{\BIBdecl}{\relax}
\BIBdecl

\bibitem{FMP3+TECS}
Y.~Takaso, H.~Oyama, and T.~Azumi, ``{Component Framework for Multiprocessor Real-Time Operating Systems},'' in \emph{Proc. of IEEE EUC}, 2022, pp. 112--119.

\bibitem{LargeScaleAndComplex}
A.~U. Alrubaee, D.~Cetinkaya, G.~A. Liebchen, and H.~Dogan, ``{A Process Model for Component-Based Model-Driven Software Development},'' \emph{Inf.}, vol.~11, p. 302, 2020.

\bibitem{TECS_ISORC2010}
T.~Azumi, H.~Takada, T.~Ukai, and H.~Oyama, ``{Wheeled Inverted Pendulum with Embedded Component System: A Case Study},'' in \emph{Proc. of IEEE ISORC}, 2010, pp. 151--155.

\bibitem{ThreeLayerCBD}
M.~Chakraborty and N.~Chaki, ``{A New Framework for Configuration Management and Compliance Checking for Component-Based Software Development},'' in \emph{Proc. of IAFOR ACSS}, 2015.

\bibitem{TECSRust}
N.~Yoshimura, H.~Oyama, and T.~Azumi, ``{TECS/Rust: Memory-safe Component Framework for Embedded Systems},'' in \emph{Proc. of IEEE ISORC}, 2024, pp. 1--11.

\bibitem{TECS}
T.~Azumi, M.~Yamamoto, Y.~Kominami, N.~Takagi, H.~Oyama, and H.~Takada, ``{A New Specification of Software Components for Embedded Systems},'' in \emph{Proc. of IEEE ISORC}, 2007, pp. 46--50.

\bibitem{Rust}
``{Rust},'' \url{https://www.rust-lang.org/}.

\bibitem{SPIKE-RT}
``{SPIKE-RT},'' \url{https://github.com/spike-rt/spike-rt}.

\bibitem{CRC}
M.~Lindner, J.~Aparicio, and P.~Lindgren, ``{C}oncurrent {R}eactive {O}bjects in {R}ust {S}ecure by {C}onstruction,'' \emph{Ada User Journal}, vol.~40, 2019.

\bibitem{ASP3+TECS}
T.~Kawada, T.~Azumi, H.~Oyama, and H.~Takada, ``Componentizing an {O}perating {S}ystem {F}eature {U}sing a {TECS} {P}lugin,'' in \emph{Proc. of IEEE CPSNA}, 2016, pp. 95--99.

\bibitem{HRMP3+TECS}
Y.~Takaso, H.~Oyama, H.~Takada, and T.~Azumi, ``{HRMP3+TECS}: {C}omponent {F}ramework for {M}ultiprocessor {R}eal-time {O}perating {S}ystem with {M}emory {P}rotection,'' in \emph{Proc. of IEEE ISORC}, 2023, pp. 86--96.

\bibitem{HRMP3+TECSv2}
Y.~Takaso, N.~Yoshimura, H.~Oyama, H.~Takada, and T.~Azumi, ``{HRMP3+TECS v2: Component Framework for Reliable Multiprocessor Real-time Operating System},'' \emph{Journal of Information Processing}, vol.~32, pp. 818--828, 2024.

\end{thebibliography}

\end{document}